# Scattering-free plasmonic Brewster effect via metasurfaces

Xinyan Zhang[1,2], Xingshuo Cui[2,3], Tong Cai[3], Weiqi Cai[4], Tony Low[5], Hongsheng Chen[1,2,6,7,*], and Xiao Lin[1,2,*]

[1]*Interdisciplinary Center for Quantum Information, State Key Laboratory of Extreme Photonics and Instrumentation, College of Information Science and Electronic Engineering, Zhejiang University, Hangzhou 310027, China.*
[2]*International Joint Innovation Center, The Electromagnetics Academy at Zhejiang University, Zhejiang University, Haining 314400, China.*
[3]*Air and Missile Defense College, Air Force Engineering University, Xi' an 710051, China.*
[4]*AVIC Research Institute for Special Structures of Aeronautical Composites, Jinan 250023, China.*
[5]*Department of Electrical and Computer Engineering, University of Minnesota, Minneapolis, Minnesota 55455, USA.*
[6]*Key Lab. of Advanced Micro/Nano Electronic Devices & Smart Systems of Zhejiang, Jinhua Institute of Zhejiang University, Zhejiang University, Jinhua 321099, China.*
[7]*Shaoxing Institute of Zhejiang University, Zhejiang University, Shaoxing 312000, China.*
[*]*Corresponding authors. Email: xiaolinzju@zju.edu.cn (X. Lin); hansomchen@zju.edu.cn (H. Chen)*

**The Brewster effect, dating back to the pioneering work of Sir David Brewster in 1815, offers a crucial route to achieve 100% energy conversion between the incident and transmitted propagating waves at an optical interface and is of fundamental importance to many practical applications, such as polarization filtering, beam steering, and optical broadband angular selectivity. However, whether the Brewster effect of surface waves can be implemented without the involvement of negative-permittivity or negative-permeability materials remains elusive. This is due to the formidable challenge to fully suppress both the parasitic scattering into propagating waves and the reflection into surface waves under the incidence of surface waves. Here, we reveal a feasible route to achieve scattering-free plasmonic Brewster effect via isotropic metasurfaces, along with the usage of positive-permittivity and positive-permeability metamaterials with both anisotropic and magnetic responses. In essence, the anisotropic response of metamaterials is judiciously designed to fully suppress the parasitic scattering into propagating waves, while the magnetic response of metamaterials facilitates the full suppression of the reflection into surface waves supported by metasurfaces. Moreover, we find that this plasmonic Brewster effect via metasurfaces can be further engineered to occur for arbitrary incident angles, giving rise to the exotic phenomenon of all-angle scattering-free plasmonic Brewster effect.**

When unpolarized light impinges on a dielectric interface at the so-called Brewster angle, the reflected light would be purely *s*-polarized. This exotic reflection phenomenon is widely known as the Brewster effect, which was discovered by Sir David Brewster in 1810's and might be the first known approach to obtaining linearly-polarized light from unpolarized one [1]. Interest in the Brewster effect in non-magnetic systems stems from the possibility that it could offer an enticing route towards achieving 100% energy conversion between the incident and transmitted *p*-polarized propagating waves [2-6]. Because of this unique capability, the Brewster effect of propagating waves is crucial to many practical applications, ranging from polarization filtering [6-9], beam steering [10-14], imaging [15,16], optical broadband angular selectivity [17-19], Brewster window in gas lasers [20,21], Brewster angle microscope [22,23], and the design of high-energy particle detectors and light sources [24,25].

Despite the longstanding research, the Brewster effect is mainly limited to propagating waves. Whether the Brewster effect can be generalized to surface waves remains under-explored [26-28]. Since surface waves (e.g. metal plasmons [29], graphene plasmons [30], spoof surface plasmons [31,32]) with high spatial confinement could mold the flow of light at the deep-subwavelength scale [33-39], the realization of plasmonic Brewster effect would have profound implications for many practical applications, such as on-chip information processing [40-42], novel guiding [43,44], and nano-imaging [45-47]. Unlike the classic Brewster effect of propagating waves, the plasmonic Brewster effect is much more complex. This is because upon the incidence of surface waves on an optical interface, there would be diversiform interconversion between evanescent waves and propagating waves [26-28,48-51]. In other words, one encounters parasitic scattering into propagating waves, in addition to the conventional reflection and transmission of surface waves. This way, the realization of plasmonic Brewster effect requires the simultaneous suppression of both the parasitic scattering into propagating waves and the reflection into surface waves.

In 2008, Ref. [49] theoretically reported an enticing route to fully suppress the parasitic scattering into propagating waves and proposed the concept of scattering-free plasmonic optics by exploiting non-magnetic anisotropic metamaterials. Notably, the reflection into surface waves in the designed non-magnetic nanostructure [49] is always nonzero for arbitrary incident angles. In 2009, Ref. [50] further extended the concept of scattering-free plasmonic optics and revealed the possibility to achieve zero reflection of surface waves by using magnetic anisotropic metamaterials. Upon close inspection of these pioneering works [49,50], the realization of scattering-free plasmonic optics relies on the usage of negative-permittivity or negative-permeability materials. In practice, the loss is generally unavoidable and relatively-large in these negative-permittivity or negative-permeability materials and would intrinsically induce parasitic scattering into propagating waves, as shown in Fig. S1. Thus, the realization of scattering-free plasmonic optics is still a long-standing scientific challenge that is highly sought after in experiments. To mitigate the loss issue, it is favorable to explore scattering-free plasmons optics in low-loss or even lossless platforms (e.g. metasurfaces at the microwave regime), without the use of negative-permittivity or negative-permeability materials. However, how to achieve the scattering-free plasmonic optics, including the scattering-free plasmonic Brewster effect, via metasurfaces has never been explored before.

Here, we find the possibility to achieve the plasmonic Brewster effect via metasurfaces, along with the usage of positive-permittivity and positive-permeability metamaterials. Under the context of scattering-free plasmonic optics via metasurfaces, we further reveal that finite magnetic response in metamaterials is a mandatory requirement to fully suppress the reflection into surface waves. This way, the revealed scattering-free plasmonic Brewster effect of *p*-polarized surface waves is fundamentally different from the conventional Brewster effect of *p*-polarized propagating waves [1,2,52], where the latter is widely known to be achievable in non-magnetic systems. Moreover, we find that this scattering-free plasmonic Brewster effect via metasurfaces could be further engineered to occur for all incident angles, leading to the first realization of an all-angle scattering-free plasmonic Brewster effect.

We begin with the conceptual illustration of scattering-free plasmonic Brewster effect via metasurfaces in Fig. 1a-f. The whole space is divided into two semi-infinite regions, i.e. the left region with $x < 0$ (denoted as region L below) and the right region with $x > 0$ (i.e. region R). In region L (R), the isotropic metasurface with a surface conductivity $\sigma_{s,L}$ ($\sigma_{s,R}$) is surrounded by a homogeneous metamaterial with a relative permittivity $\bar{\bar{\varepsilon}}_L = \text{diag}\,[\varepsilon_{L,\parallel}, \varepsilon_{L,\parallel}, \varepsilon_{L,\perp}]$ ($\bar{\bar{\varepsilon}}_R = \text{diag}\,[\varepsilon_{R,\parallel}, \varepsilon_{R,\parallel}, \varepsilon_{R,\perp}]$) and a relative permeability $\bar{\bar{\mu}}_L = \text{diag}\,[\mu_{L,\parallel}, \mu_{L,\parallel}, \mu_{L,\perp}]$ ($\bar{\bar{\mu}}_R = \text{diag}\,[\mu_{R,\parallel}, \mu_{R,\parallel}, \mu_{R,\perp}]$). Under this scenario, the dispersion of $p$-polarized surface waves [29,52] can be obtained as

$$k_{j,z} = -\frac{2\omega\varepsilon_0 \varepsilon_{j,\parallel}}{\sigma_{s,j}} \tag{1}$$

$$|\bar{k}_{j,\parallel}|^2 = \frac{\varepsilon_{j,\perp}}{\varepsilon_{j,\parallel}}(-k_{j,z}^2 + k_0^2 \varepsilon_{j,\parallel}\mu_{j,\parallel}) \tag{2}$$

where the subscript j represents L or R, $\bar{k}_j = \bar{k}_{j,\parallel} \pm \hat{z} k_{j,z}$ is the wavevector of $p$-polarized surface waves, $\bar{k}_{j,\parallel} = \hat{x} k_{j,x} + \hat{y} k_{j,y}$, $\varepsilon_0$ is the permittivity of free space, $k_0 = \omega/c$, $\omega$ is the angular frequency, and $c$ is the speed of light in vacuum. For conceptual illustration, all material loss is neglected below, and meanwhile, $\varepsilon_{L,\parallel}$, $\varepsilon_{L,\perp}$, $\varepsilon_{R,\parallel}$, and $\varepsilon_{R,\perp}$ are chosen to be positive real numbers. Since $k_{j,z}$ is purely imaginary for surface waves, the surface conductivity $\sigma_{s,j}$ should be purely imaginary with $\text{Im}(\sigma_{s,j}) > 0$.

When there is no parasitic scattering into propagating waves, the magnetic fields (i.e. $\bar{H}_i$, $\bar{H}_r$ and $\bar{H}_t$) of incident, reflected, and transmitted $p$-polarized surface waves can be assumed as

$$\bar{H}_i = \hat{z} \times \hat{k}_{i,\parallel} H_0 e^{i\bar{k}_{i,\parallel}\bar{r}_{i,\parallel} \pm ik_{i,z}z} \tag{3}$$

$$\bar{H}_r = \hat{z} \times \hat{k}_{r,\parallel} H_0 e^{i\bar{k}_{r,\parallel}\bar{r}_{r,\parallel} \pm ik_{r,z}z} \cdot \text{r}_{sw} \tag{4}$$

$$\bar{H}_t = \hat{z} \times \hat{k}_{t,\parallel} H_0 e^{i\bar{k}_{t,\parallel}\bar{r}_{t,\parallel} \pm ik_{t,z}z} \cdot \text{t}_{sw} \tag{5}$$

where $r_{sw}$ and $t_{sw}$ are the reflection and transmission coefficients, respectively; $\bar{k}_i = \bar{k}_{i,\parallel} \pm \hat{z} k_{i,z}$ and $\bar{k}_{i,\parallel} =$

$\hat{x}k_{i,x} + \hat{y}k_{i,y}$; $\bar{k}_r = \bar{k}_{r,\parallel} \pm \hat{z}k_{r,z}$ and $\bar{k}_{r,\parallel} = \hat{x}k_{r,x} + \hat{y}k_{r,y}$; $\bar{k}_t = \bar{k}_{t,\parallel} \pm \hat{z}k_{t,z}$ and $\bar{k}_{t,\parallel} = \hat{x}k_{t,x} + \hat{y}k_{t,y}$; $\hat{k}_{i,\parallel} = \bar{k}_{i,\parallel}/|\bar{k}_{i,\parallel}|$, $\hat{k}_{r,\parallel} = \bar{k}_{r,\parallel}/|\bar{k}_{r,\parallel}|$, and $\hat{k}_{t,\parallel} = \bar{k}_{t,\parallel}/|\bar{k}_{t,\parallel}|$; $H_0$ is a constant; in the superscripts, $+$ ($-$) is used for the region with $z > 0$ ($z < 0$). Without loss of generality, we let the surface waves be incident from region L. Under this scenario, we have $k_{i,z} = k_{r,z} = k_{L,z}$, $k_{t,z} = k_{R,z}$, $|\bar{k}_{i,\parallel}| = |\bar{k}_{r,\parallel}| = |\bar{k}_{L,\parallel}|$, and $|\bar{k}_{t,\parallel}| = |\bar{k}_{R,\parallel}|$. Meanwhile, the related electric fields (i.e. $\bar{E}_i$, $\bar{E}_r$ and $\bar{E}_t$) for the incident, reflected and transmitted surface waves can be obtained by using Faraday's law ($\nabla \times \bar{E} = i\omega\bar{\mu}\bar{H}$). From the boundary conditions at the plane of $x = 0$, namely $-\hat{x} \times (\bar{E}_i + \bar{E}_r - \bar{E}_t) = 0$ and $-\hat{x} \times (\bar{H}_i + \bar{H}_r - \bar{H}_t) = 0$, we further have

$$k_{i,z} = k_{r,z} = k_{L,z} = k_{t,z} = k_{R,z} \tag{6}$$

$$k_{i,y} = k_{r,y} = k_{t,y} \tag{7}$$

$$\cos\theta_i(1 - r_{sw}) - \cos\theta_t t_{sw} = 0 \tag{8}$$

$$\frac{\sin\theta_i}{\varepsilon_{L,\parallel}}(1 + r_{sw}) - \frac{\sin\theta_t}{\varepsilon_{R,\parallel}} t_{sw} = 0 \tag{9}$$

$$\frac{|\bar{k}_{L,\parallel}|}{\varepsilon_{L,\perp}}(1 + r_{sw}) - \frac{|\bar{k}_{R,\parallel}|}{\varepsilon_{R,\perp}} t_{sw} = 0 \tag{10}$$

where $\theta_i$ and $\theta_t$ are the incident and transmitted angles. According to the geometry optics, we have $\sin\theta_i = k_{i,y}/|\bar{k}_{L,\parallel}|$ and $\sin\theta_t = k_{t,y}/|\bar{k}_{R,\parallel}|$.

Upon close inspections, equation (6) is intrinsically related the modal matching condition. That is, the profile of surface waves along the $z$ direction would remain unchanged across the boundary at the plane of $x = 0$. From equations (1 & 6), the scattering-free plasmonic optics via metasurfaces actually requires that

$$\frac{\varepsilon_{L,\parallel}}{\sigma_{s,L}} = \frac{\varepsilon_{R,\parallel}}{\sigma_{s,R}} \tag{11}$$

In other words, when the modal matching condition or equation (11) is not fulfilled, there would always be the emergence of parasitic scattering into propagating waves, as shown in Fig. 1c&f.

On the other hand, in order to solve the unknown parameters of $r_{sw}$ and $t_{sw}$ in equations (8-10), an extra constraint should be applied to equations (9-10), namely

$$\frac{|\bar{k}_{L,\|}|}{\varepsilon_{L,\perp}} \cdot \frac{\sin\theta_t}{\varepsilon_{R,\|}} = \frac{|\bar{k}_{R,\|}|}{\varepsilon_{R,\perp}} \frac{\sin\theta_i}{\varepsilon_{L,\|}} \tag{12}$$

According to equations (2, 6, 7 & 12), the scattering-free plasmonic optics via metasurfaces additionally requires that

$$\varepsilon_{L,\|}\mu_{L,\|} = \varepsilon_{R,\|}\mu_{R,\|} \tag{13}$$

Moreover, according to equations (8-10 & 12), the reflection coefficient can be obtained as

$$r_{sw} = \left(1 - \frac{|\bar{k}_{R,\|}|\cos\theta_t}{|\bar{k}_{L,\|}|\cos\theta_i}\frac{\varepsilon_{R,\|}}{\varepsilon_{L,\|}}\right) \bigg/ \left(1 + \frac{|\bar{k}_{R,\|}|\cos\theta_t}{|\bar{k}_{L,\|}|\cos\theta_i}\frac{\varepsilon_{R,\|}}{\varepsilon_{R,\|}}\right) \tag{14}$$

Particularly, when $r_{sw} = 0$ in equation (14), the scattering-free plasmonic Brewster effect requires that

$$\frac{|\bar{k}_{L,\|}|\cos\theta_i}{|\bar{k}_{R,\|}|\cos\theta_t} = \frac{\varepsilon_{R,\|}}{\varepsilon_{L,\|}} \tag{15}$$

We highlight that the scattering-free plasmonic Brewster effect via metasurfaces cannot be achieved in systems with $\bar{\bar{\mu}}_L = \bar{\bar{\mu}}_R$, including the non-magnetic systems with $\bar{\bar{\mu}}_L = \bar{\bar{\mu}}_R = 1 \cdot \bar{\bar{I}}$ as shown in Fig. 1b&e. For example, when $\bar{\bar{\mu}}_L = \bar{\bar{\mu}}_R = 1 \cdot \bar{\bar{I}}$ in Fig. 1b&e, equations (13 & 15) indicate that $\varepsilon_{L,\|} = \varepsilon_{R,\|}$ and $|\bar{k}_{L,\|}|\cos\theta_i = |\bar{k}_{R,\|}|\cos\theta_t$. Moreover, from the momentum matching condition as governed by equation (7), we further have $|\bar{k}_{L,\|}|\sin\theta_i = |\bar{k}_{R,\|}|\sin\theta_t$. This way, we always have $|\bar{k}_{L,\|}| = |\bar{k}_{R,\|}|$, which further leads to $\sigma_{s,L} = \sigma_{s,R}$ and $\varepsilon_{L,\perp} = \varepsilon_{R,\perp}$. Since $\sigma_{s,L} = \sigma_{s,R}$, $\bar{\bar{\varepsilon}}_L = \bar{\bar{\varepsilon}}_R$ and $\bar{\bar{\mu}}_L = \bar{\bar{\mu}}_R$, regions L and R are actually the same, and there is essentially the disappearance of electromagnetic boundary at the plane of $x = 0$. This way, under the scenario of $\bar{\bar{\mu}}_L = \bar{\bar{\mu}}_R$ and that regions L and R are not the same, the scattering-free plasmonic Brewster effect via metasurfaces cannot be achieved, and there is always nonzero reflection, as shown in Fig. 1b&e.

According to above analyses and equations (11, 13 &15), the scattering-free plasmonic Brewster effect via metasurfaces requires the simultaneous satisfaction of the following conditions, namely

$$\bar{\bar{\mu}}_L \neq \bar{\bar{\mu}}_R, \frac{\varepsilon_{L,\|}}{\sigma_{s,L}} = \frac{\varepsilon_{R,\|}}{\sigma_{s,R}}, \text{ and } \varepsilon_{L,\|}\mu_{L,\|} = \varepsilon_{R,\|}\mu_{R,\|} \tag{16}$$

In other words, the existence of magnetic response for the surrounding metamaterials is a mandatory condition to achieve the scattering-free plasmonic Brewster effect. Upon close inspections, equation (16) can be fulfilled if all parameters of $\varepsilon_{L,\parallel}$, $\varepsilon_{L,\perp}$, $\varepsilon_{R,\parallel}$, $\varepsilon_{R,\perp}$, $\mu_{L,\parallel}$, $\mu_{L,\perp}$, $\mu_{R,\parallel}$, and $\mu_{R,\perp}$ are positive. For conceptual demonstration, we show in Fig. 1a&d the emergence of scattering-free plasmonic Brewster effect when the positive-permittivity and positive-permeability metamaterials are used.

Under the scenario of $\bar{\bar{\mu}}_L \neq \bar{\bar{\mu}}_R$, according to equations (7 & 15), the plasmonic Brewster angle $\theta_B$ can be further obtained as

$$\theta_B = \arctan\left(\sqrt{\frac{|\bar{k}_{R,\parallel}|^2 \varepsilon_{R,\parallel}^2 - |\bar{k}_{L,\parallel}|^2 \varepsilon_{L,\parallel}^2}{\varepsilon_{R,\parallel}^2 - \varepsilon_{L,\parallel}^2}}\Big/|\bar{k}_{L,\parallel}|\right) \qquad (17)$$

Figure 2 quantitatively shows the reflection $|r_{sw}|$ of surface waves as a function of the incident angle $\theta_i$. The zero reflection is achievable only at the plasmonic Brewster angle (i.e. $\theta_i = \theta_B$) under the scenario of $\bar{\bar{\mu}}_L \neq \bar{\bar{\mu}}_R$. For illustration, $|\bar{k}_{L,\parallel}| = 10.58 k_0$ and $|\bar{k}_{R,\parallel}| = 15.87 k_0$ are designed in Fig. 2a, and accordingly, we have $\theta_B = 66.67^o$. This phenomenon is further numerically verified in Fig. 2b-d through the field simulation via the commercial software (e.g. COMSOL simulation). To be specific, there are always reflected surface waves no matter $\theta_i < \theta_B$ in Fig. 2b or $\theta_i > \theta_B$ in Fig. 2d but zero reflected surface waves if $\theta_i = \theta_B$ in Fig. 2c.

Figure 3 shows that the vertical component of relative permittivity $\varepsilon_{j,\perp}$ could offer an extra degree of freedom to flexibly tailor the plasmonic Brewster angle. The underlying reason is that as governed by equation (16), the conditions for scattering-free plasmonic Brewster effect have no specific requirement on the choice of $\varepsilon_{j,\perp}$. For illustration, Fig. 3a shows $\theta_{i,B}$ and $\theta_{rt,B}$ as a function of $\varepsilon_{R,\perp}$, where $\theta_{i,B} = \theta_B$ is the incident plasmonic Brewster angle and $\theta_{rt,B}$ corresponds to the angle between between $\bar{k}_{r,\parallel}$ and $\bar{k}_{t,\parallel}$. Through the judicious design of $\varepsilon_{R,\perp}$, we find that $\theta_{i,B}$ can vary from $0^o$ to $90^o$. Particularly, Fig. 3a shows the possibility to achieve the scattering-free plasmonic Brewster effect under the normal incidence with $\theta_i = \theta_{i,B} = 0^o$. By contrast, the conventional Brewster angle of p-polarized propagating waves [1,2] is generally non-zero in non-magnetic and isotropic systems. On the other hand, through the judicious design of $\varepsilon_{R,\perp}$, we further find that $\theta_{rt,B}$ can change from $0^o$ to $180^o$. That is, the scattering-free plasmonic Brewster effect can have $\theta_{rt,B} < 90^o$ in Fig. 2b, $\theta_{rt,B} = 90^o$ in Fig. 2c, or $\theta_{rt,B} > 90^o$ in Fig. 2d, as facilitated by the existence of magnetic and anisotropic responses for surrounding metamaterials. As background, since the wavevectors of reflected and transmitted *p*-polarized propagating waves are perpendicular to each other in non-magnetic and isotropic systems, the corresponding $\theta_{rt,B}$ for conventional Brewster effects of *p*-polarized propagating waves would be exactly $90^o$ [1,2].

Last but not the least, we find that the reflection coefficient in equation (14) could be zero for arbitrary incident angles in Fig. 4a-b, if $\mu_{L,\parallel} = \mu_{R,\parallel}$ is additionally added to the conditions of scattering-free plasmonic Brewster effect in equation (16). Briefly speaking, the exotic phenomenon of all-angle scattering-free plasmonic Brewster effect would emerge, as shown in Fig. 4c-d, if the following conditions are simultaneously fulfilled, namely

$$\mu_{L,\perp} \neq \mu_{R,\perp}, \mu_{L,\parallel} = \mu_{R,\parallel}, \bar{\bar{\varepsilon}}_L = \bar{\bar{\varepsilon}}_R, \text{ and } \sigma_{s,L} = \sigma_{s,R} \quad (18)$$

To be specific, by substituting equation (18) into equation (2), we always have $|\bar{k}_{L,\parallel}| = |\bar{k}_{R,\parallel}|$. Since $|\bar{k}_{L,\parallel}| = |\bar{k}_{R,\parallel}|$ and $\bar{\bar{\varepsilon}}_L = \bar{\bar{\varepsilon}}_R$, the zero-reflection condition in equation (15) further requires that $\theta_i = \theta_t$. As a result, both the propagation direction (i.e. $\theta_i = \theta_t$) and the wavelength (i.e. $2\pi/|\bar{k}_{L,\parallel}| = 2\pi/|\bar{k}_{R,\parallel}|$) of surface waves would keep unchanged across the boundary at the plane of $x = 0$ for arbitrary incident angles. Accordingly, the boundary between regions L and R at the plane of $x = 0$ would have no disturbance to the propagation of surface waves excited by a dipolar source as shown in Fig. 4c.

In conclusion, we have revealed a feasible paradigm to realize the scattering-free plasmonic optics, including the scattering-free plasmonic Brewster effect, without the involvement of negative-permittivity and negative-permeability metamaterials by readily exploiting metasurfaces. Moreover, we have found that the usage of metamaterials with anisotropic and magnetic responses could offer a powerful capability to tailor the plasmonic Brewster angle, such as the possibility to achieve the all-angle scattering-free plasmonic Brewster effect. These findings might provide a useful theoretical guidance for the possible observation of scattering-free plasmonic optics in the near future and may trigger the continual exploration of other exotic photonic phenomena in the platform of scattering-free plasmonic optics, such as superscattering [53-56] of surface waves and Kerker scattering [57-59] of surface waves.

**ACKNOWLEDGMENTS**
X.L. acknowledges the support partly from the Fundamental Research Funds for the Central Universities under Grant No. 226-2024-00022, the National Natural Science Fund for Excellent Young Scientists Fund Program (Overseas) of China, the National Natural Science Foundation of China (NSFC) under Grant No. 62175212, and Zhejiang Provincial Natural Science Fund Key Project under Grant No. LZ23F050003. and. H.C. acknowledges the support from the Key Research and Development Program of the Ministry of Science and Technology under Grants No. 2022YFA1404704, 2022YFA1405200, and 2022YFA1404902, the National Natural Science Foundation of China (NNSFC) under Grants No. 61975176, the Key Research and Development Program of Zhejiang Province under Grant No.2022C01036, and the Fundamental Research Funds for the Central Universities.

**DATA AVAILABILITY**
The data that support the findings of this study are available within the article and its supplementary material.

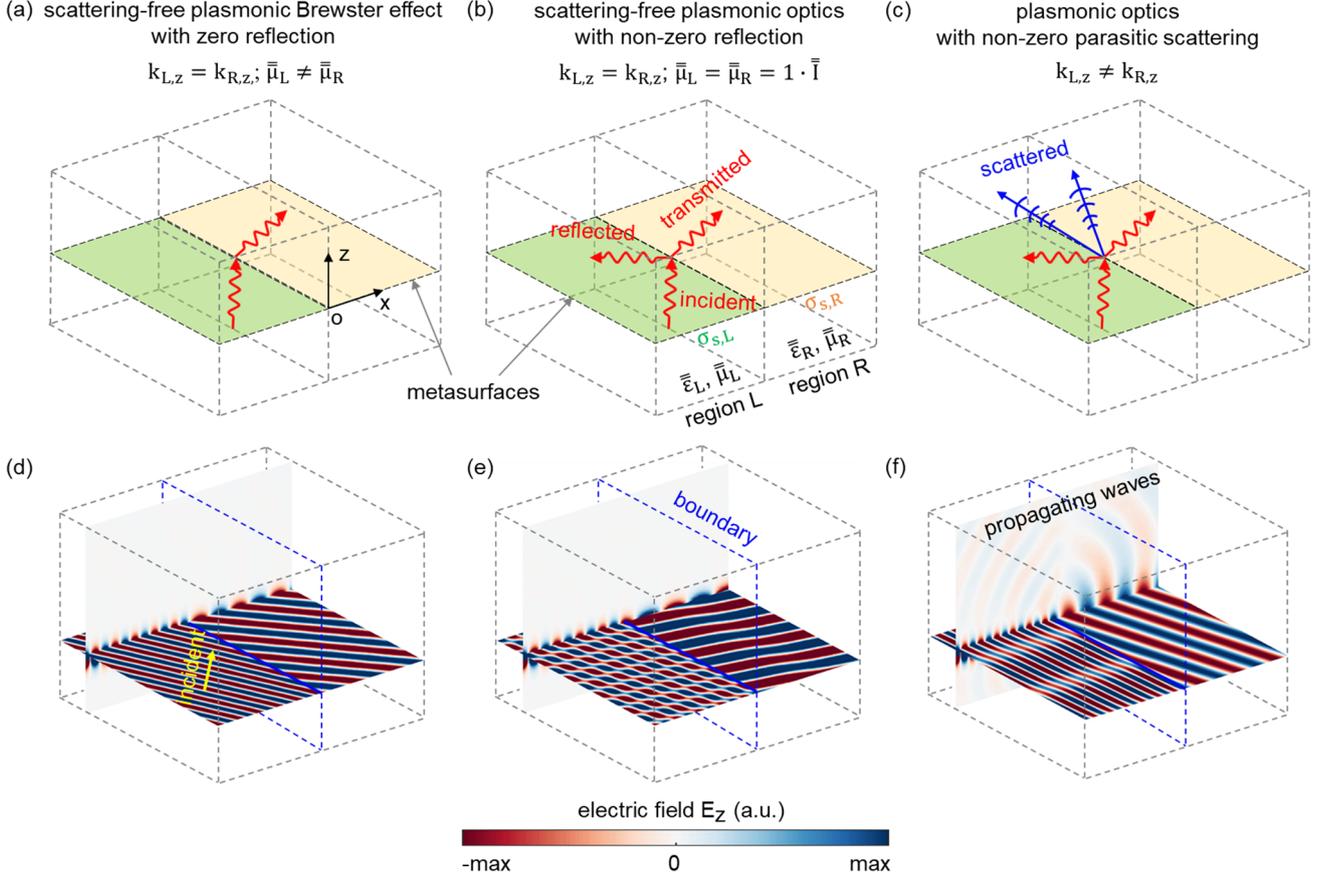

FIG. 1. Conceptual illustration of scattering-free plasmonic Brewster effect via metasurfaces for $p$-polarized surface waves. The left (right) metasurface with a surface conductivity $\sigma_{s,L}$ ($\sigma_{s,R}$) is surrounded by a homogeneous metamaterial with a relative permittivity $\bar{\bar{\varepsilon}}_L$ ($\bar{\bar{\varepsilon}}_R$) and a relative permeability $\bar{\bar{\mu}}_L$ ($\bar{\bar{\mu}}_R$). (a, d) Scattering-free plasmonic Brewster effect with zero reflection. For illustration, the following structural setup is used in (d): $\sigma_{s,L} = 2.5 \times 10^{-3}$i S, $\bar{\bar{\varepsilon}}_L = \text{diag}[1, 1, 1]$, $\bar{\bar{\mu}}_L = \text{diag}[2, 2, 1]$, $\sigma_{s,R} = 5 \times 10^{-3}$i S, $\bar{\bar{\varepsilon}}_R = \text{diag}[2, 2, 1]$ and $\bar{\bar{\mu}}_R = \text{diag}[1, 1, 1]$. (b, e) Scattering-free plasmonic optics with non-zero reflection. $\sigma_{s,L} = 2.5 \times 10^{-3}$i S, $\bar{\bar{\varepsilon}}_L = \text{diag}[1, 1, 1.18]$, $\bar{\bar{\mu}}_L = \text{diag}[1, 1, 1]$, $\sigma_{s,R} = 2.5 \times 10^{-3}$i S, $\bar{\bar{\varepsilon}}_R = \text{diag}[1, 1, 0.7]$ and $\bar{\bar{\mu}}_R = \text{diag}[1, 1, 1]$ are used in (e). (c, f) Plasmonic optics with non-zero parasitic scattering and non-zero reflection. $\sigma_{s,L} = 2.5 \times 10^{-3}$i S, $\bar{\bar{\varepsilon}}_L = \text{diag}[1, 1, 1]$, $\bar{\bar{\mu}}_L = \text{diag}[2, 2, 1]$, $\sigma_{s,R} = 9 \times 10^{-3}$i S, $\bar{\bar{\varepsilon}}_R = \text{diag}[1, 1, 1]$ and $\bar{\bar{\mu}}_R = \text{diag}[1, 1, 1]$ are used in (f).

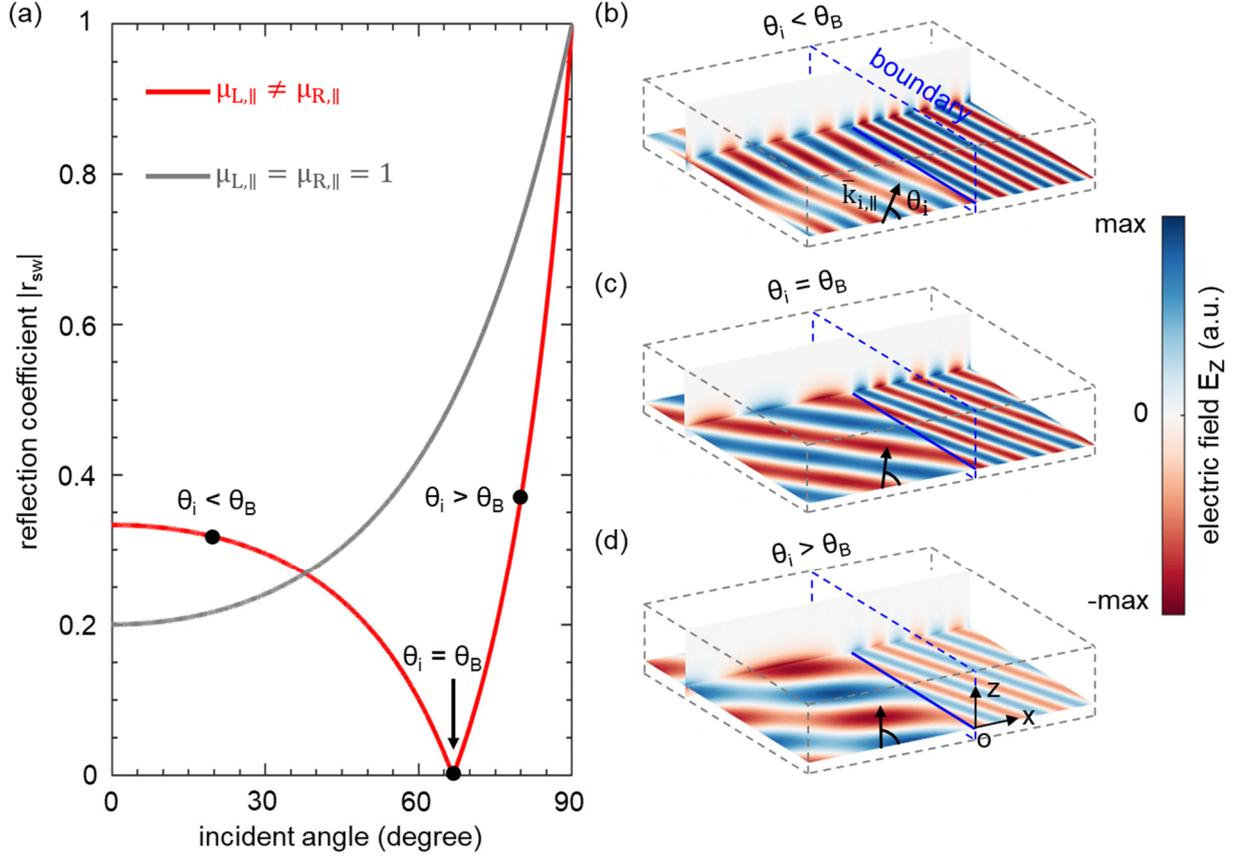

FIG. 2. Zero reflection for scattering-free plasmonic Brewster effect of *p*-polarized surface waves via metasurfaces. (a) Reflection coefficient $r_{sw}$ as a function of the incident angle. The red line corresponds to the magnetic case with $\mu_{L,\parallel} \neq \mu_{R,\parallel}$, where $\sigma_{s,L} = 3.18 \times 10^{-3}$ i S, $\bar{\bar{\varepsilon}}_L = \text{diag}[3,3,12]$, $\bar{\bar{\mu}}_L = \text{diag}[1,1,1]$, $\sigma_{s,R} = 1.06 \times 10^{-3}$ i S, $\bar{\bar{\varepsilon}}_R = \text{diag}[1,1,9]$, and $\bar{\bar{\mu}}_R = \text{diag}[3,3,1]$ are used. The gray line corresponds to the non-magnetic case with $\mu_{L,\parallel} = \mu_{R,\parallel} = 1$, where $\sigma_{s,R} = 1.06 \times 10^{-3}$ i S, $\bar{\bar{\varepsilon}}_L = \text{diag}[1,1,4.31]$, $\bar{\bar{\mu}}_L = \text{diag}[1,1,1]$, $\sigma_{s,R} = 1.06 \times 10^{-3}$ i S, $\bar{\bar{\varepsilon}}_R = \text{diag}[1,1,9.69]$, and $\bar{\bar{\mu}}_R = \text{diag}[1,1,1]$ are used. (b-d) Distribution of the electric field $E_z$ for various incident angles via COMSOL simulation. The structural setup in (b-d) is the same as that of the magnetic case in (a).

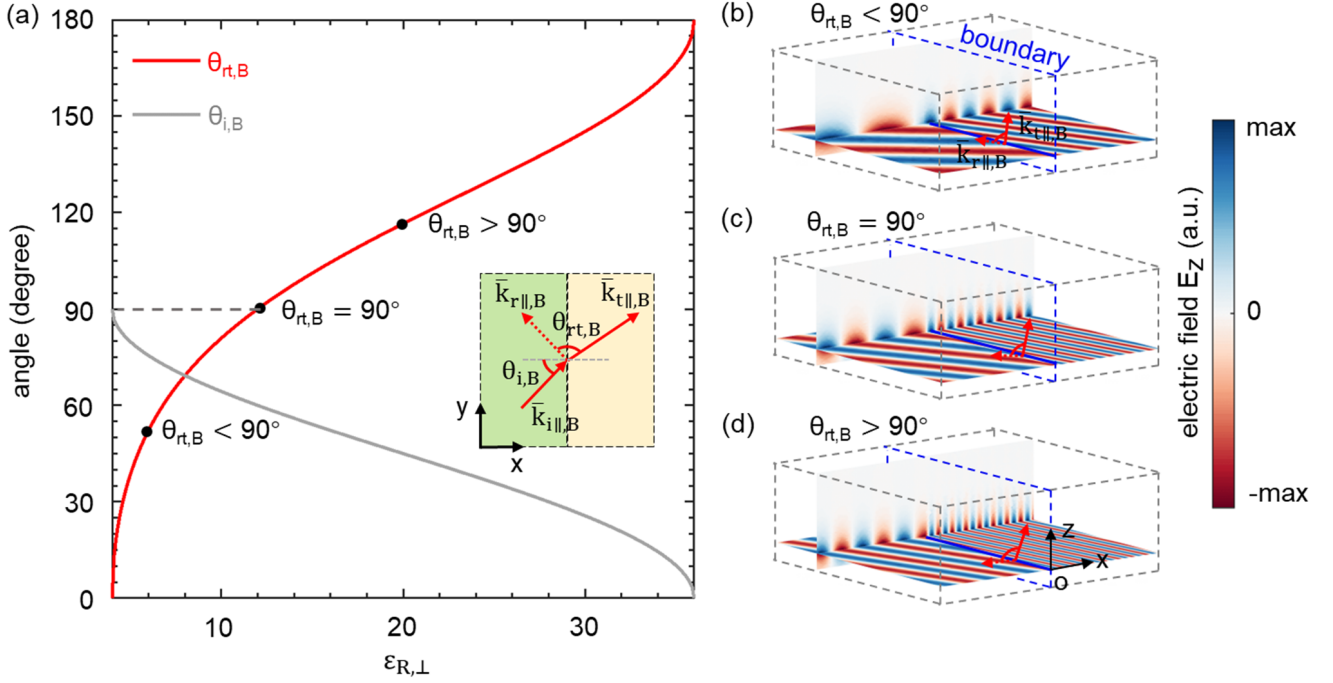

FIG. 3. Plasmonic Brewster angle. The structural setup is the same as Fig. 2(b-d), except for $\varepsilon_{R,\perp}$. (a) Dependence of $\theta_{i,B}$ and $\theta_{rt,B}$ on $\varepsilon_{R,\perp}$, where $\theta_{i,B}$ is the incident plasmonic Brewster angle and $\theta_{rt,B}$ is the angle between the reflected wavevector $\bar{k}_{r,\parallel}$ and the transmitted wavevector $\bar{k}_{t,\parallel}$. (b-d) Distribution of the electric field $E_z$ for various $\theta_{rt,B}$. For illustration, we set $\varepsilon_{R,\perp} = 6$ for the case of $\theta_{rt,B} < 90°$ in (b), $\varepsilon_{R,\perp} = 12$ for the case of $\theta_{rt,B} = 90°$ in (c), and $\varepsilon_{R,\perp} = 20$ for the case of $\theta_{rt,B} > 90°$ in (d).

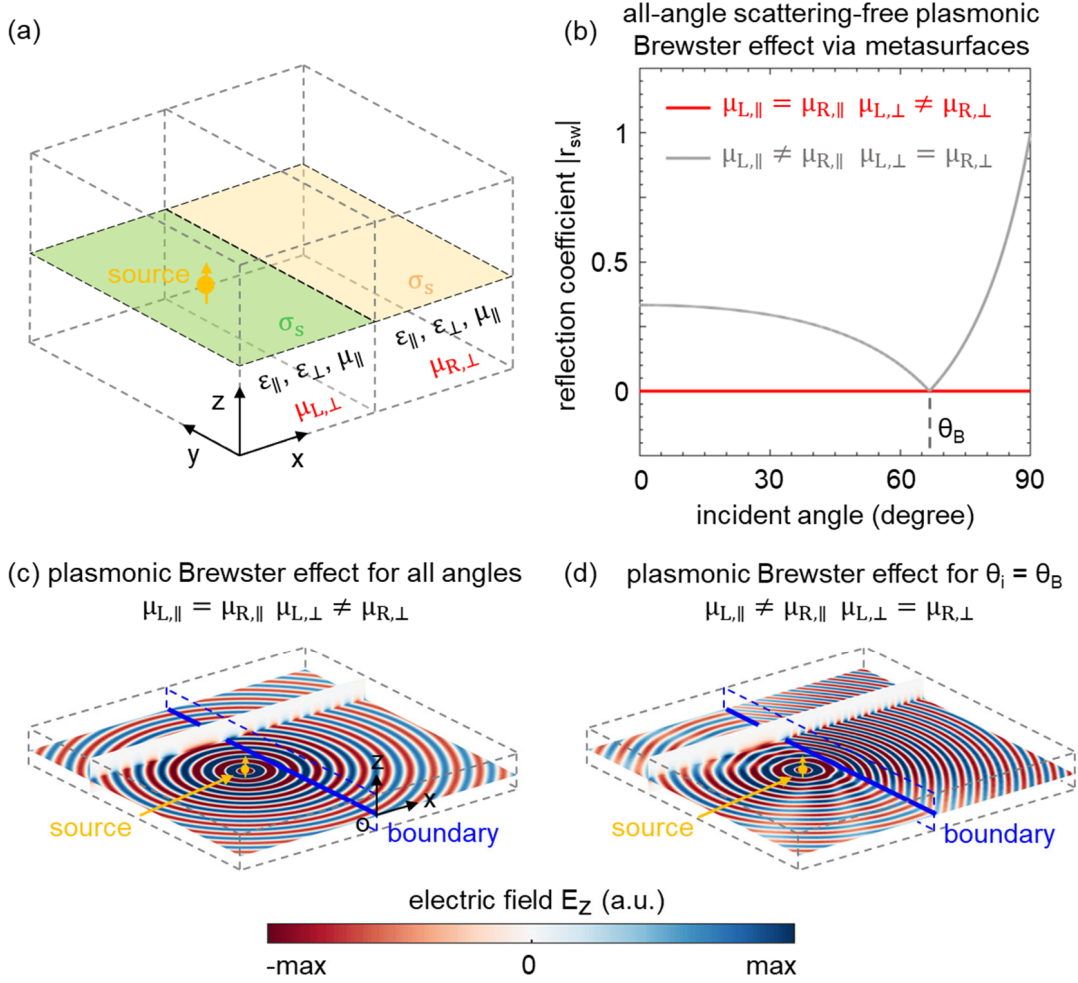

FIG. 4. All-angle scattering-free plasmonic Brewster effect of *p*-polarized surface waves via metasurfaces. (a) Structural schematic. (b) Reflection coefficient as a function of the incident angle. (c, d) Field distribution of *p*-polarized surface waves excited by a dipolar source with its dipole moment along the $\hat{z}$ direction. When the conditions of $\sigma_{s,L} = \sigma_{s,R}$, $\bar{\bar{\varepsilon}}_L = \bar{\bar{\varepsilon}}_R$, $\mu_{L,\parallel} = \mu_{R,\parallel}$, and $\mu_{L,\perp} \neq \mu_{R,\perp}$ are fulfilled, the all-angle scattering-free plasmonic Brewster effect could occur. For illustration, $\sigma_{s,L} = 3.18 \times 10^{-3}$i S, $\bar{\bar{\varepsilon}}_L = \text{diag}[3, 3, 12]$, $\bar{\bar{\mu}}_L = \text{diag}[1, 1, 1]$, $\sigma_{s,R} = 3.18 \times 10^{-3}$i S, $\bar{\bar{\varepsilon}}_R = \text{diag}[3, 3, 12]$ and $\bar{\bar{\mu}}_R = \text{diag}[1, 1, 5]$ are used in (b, c). For comparison, the conventional single-angle scattering-free plasmonic Brewster effect is also shown in (b, d), along with the usage of $\sigma_{s,L} = 3.18 \times 10^{-3}$i S, $\bar{\bar{\varepsilon}}_L = \text{diag}[3, 3, 12]$, $\bar{\bar{\mu}}_L = \text{diag}[1, 1, 1]$, $\sigma_{s,R} = 1.06 \times 10^{-3}$i S, $\bar{\bar{\varepsilon}}_R = \text{diag}[1, 1, 9]$ and $\bar{\bar{\mu}}_R = \text{diag}[3, 3, 1]$.